\documentclass[12pt]{iopart} 

\begin{document}

\title{Comment}{Comment on 'Note on the dog-and-rabbit chase
problem in introductory kinematics'}

\author{Z.~K.~Silagadze$^{1,2}$ and G.~I.~Tarantsev$^2$}

\address{$^1$ Budker Institute of Nuclear Physics, 630 090, Novosibirsk, 
Russia}

\address{$^2$ Department of physics, Novosibirsk State University, 630 090,
Novosibirsk, Russia}

\ead{silagadze@inp.nsk.su}

\begin{abstract}
We comment on the recent paper by Yuan Qing-Xin and Du Yin-Xiao 
(2008 {\it Eur.\ J.\ Phys.\ } {\bf 29} N43--N45).
\end{abstract}

\maketitle 

In a recent interesting letter \cite{1} Yuan Qing-Xin and Du Yin-Xiao
presented  a new simple derivation of the critical region for the 
dog-and-rabbit chase problem. We would like to indicate that the
equation (6) of \cite{1}, on which their treatment is based, can be 
obtained in a very simple and transparent way.

Let  radius-vectors of the rabbit and the dog are $\vec{r}_1$ and 
$\vec{r}_2$, respectively, and the corresponding velocities  $\vec{V}_1$ and 
$\vec{V}_2$. Relative radius-vector $\vec{r}=\vec{r}_1-\vec{r}_2$ is parallel 
to the dog's velocity $\vec{V}_2$ because  the dog is always heading towards 
the rabbit. Hence it is perpendicular to the dog's acceleration 
$\dot{\vec{V}}_2$ as the dog runs with the constant in magnitude velocity and 
therefore $\vec{V}_2\cdot \dot{\vec{V}}_2=0$. Using this fact and taking 
into account that $\dot{\vec{V}}_1=0$, we get easily 
\begin{equation}
\frac{d}{dt}\left [\vec{r}\cdot (\vec{V}_1+\vec{V}_2)\right]=
(\vec{V}_1-\vec{V}_2)\cdot (\vec{V}_1+\vec{V}_2)=V_1^2-V_2^2.
\label{eq1}
\end{equation}
But r.h.s of this equation is a constant and if we integrate both sides of 
it with respect to time from $t=0$ to $t=T$, when the dog 
catches the rabbit, and solve with respect to the duration $T$ of the chase,
we get
\begin{equation}
T=\frac{\left .\left [\vec{r}\cdot (\vec{V}_1+\vec{V}_2)\right]\right |_
{t=0}}{V_2^2-V_1^2}.
\label{eq2}
\end{equation}
A simple glance on the figure 1 from \cite{1} is sufficient to deduce that 
at the beginning of the chase
$$\left . \left [\vec{r}\cdot (\vec{V}_1+\vec{V}_2)\right] \right |_{t=0}=
L\left ( V_1\sin{\alpha}+V_2 \right ).$$
Therefore, for the distance $s=V_1T$, run by the rabbit before being caught 
by the dog, we get 
\begin{equation}
s=L\frac{V_1\left ( V_1\sin{\alpha}+V_2 \right )}{V_2^2-V_1^2}=
L\frac{e\left (e\sin{\alpha}+1\right )}{1-e^2},
\label{eq3}
\end{equation} 
with $e=V_1/V_2$. This is just the equation (6) from \cite{1}.


\section*{References}
\begin{thebibliography}{99}
\bibitem{1}
Qing-Xin  Y and Yin-Xiao D 2008 
Note on the dog-and-rabbit chase problem in introductory kinematics
{\it Eur.\ J.\ Phys.\ } {\bf 29} N43--N45
\end{thebibliography}
\end{document}